\documentclass[aps,prl,floatfix,twocolumn,footinbib,amsmath,amssymb]{revtex4}
\usepackage{amsmath,amsthm,amssymb,color,psfrag}
\usepackage{url}
\usepackage{latexsym}
\usepackage{graphicx}
\usepackage{slashed}

\definecolor{darkred}{rgb}{0.6,0.0,0.0}
\definecolor{darkblue}{rgb}{0.0,0.0,0.5}
\definecolor{darkgreen}{rgb}{0.0,0.5,0.0}
\definecolor{brown}{rgb}{0.0,0.0,0.0}

\newcommand{\rap}{y}

\begin{document}


\title{Quit Using Pseudorapidity, Transverse Energy, and Massless Constituents}
\author{Jason Gallicchio$^{a}$ and Yang-Ting Chien$^{b}$}
\affiliation{
$^a$ Department of Physics, Harvey Mudd College, 301 Platt Blvd., Claremont, CA 91711, USA\\
$^b$ Center for Theoretical Physics, Massachusetts Institute of Technology, Cambridge, MA 02139, USA
}

\begin{abstract}
Use a massive jet's true-rapidity instead of its pseudorapidity,
even for 
event displays and for determining if a jet is near a calorimeter edge.
Use 
transverse momentum instead of transverse energy since only the former is conserved.
Use massive constituents because using massless constituents reduces the jet
mass by an amount proportional to the square of the number of hadrons in the jet,
and can amount to several GeV.
These three recommendations are important for precision measurements when jets are constructed by adding constituent 4-vectors.
\end{abstract}
\maketitle

\section{Quit Using Pseudorapidity}
The geometrically defined pseudorapidity ($\eta$) of a modern
(e.g. anti-$k_T$ \cite{Cacciari:2008gp}) jet does
\emph{not} really correspond to ``the $\eta$ location
in the detector where the jet's $p_T$ is concentrated.''
This is better captured by the true-rapidity ($y$) of the jet.
True-rapidity is even the right cut to use to avoid the
edges or cracks of the calorimeter.


The days of massless cone jets are over,
and $\eta$ no longer fits the intuition it used to.
LHC jets are made by
summing 4-vectors of ``constituents.''  These constituents can be
truth-hadrons, calorimeter cells, tracks, or particle-flow \cite{particleflow} objects.
We will treat these constituents as massless until the last section,
so here their pseudorapidity will be equal to their true-rapidity.
However, once you add massless 4-vectors together
to make a jet, the jet becomes massive and its $\eta$ differs from $y$.
%
%
%
Here we hope to give you some intuition on the differences,
and convince you to always use true-rapidity.
Intuition for the differences is not easy since they become identical
in the two simplest limits:  when the jet 4-vector becomes massless,
and also when it becomes purely transverse (perpendicular to the beam).

The following definitions of true-rapidity and pseudorapidity are
standard at hadronic colliders and used in FastJet \cite{Cacciari:2011ma}:
%
\[
\arraycolsep=15pt
\begin{array}{l|l}
\rap \textrm{ (true-rapidity) }  &
\eta \textrm{ (pseudo-rapidity) } \\[5pt]
\displaystyle = \tanh^{-1} \frac{p_z}{E}  &
\displaystyle = \tanh^{-1} \frac{p_z}{ |\vec p| }  \\[10pt]
\displaystyle = \frac{1}{2} \ln \frac{ E + p_z } { E - p_z } &
\displaystyle = - \ln \left( \tan \frac{\theta}{2} \right)
\end{array}
\]




\vspace{1em}

\subsection{For Massive 4-vectors, $|\eta| > |y|$}

From the first definition, the fact that $\tanh^{-1}$
increases monotonically, and $E > |\vec p|$ for massive 4-vectors, it follows that $|\eta| > |y|$.
%
To remember the direction of the inequality,
remember $\eta$ blows up when the 3-vector points along the z-axis,
while $y$ stays finite for massive 4-vectors.


The figure below shows two `jets', each composed of two massless
constituents with equal $p_T$, whose location is plotted as colored squares.
The $+$ marks the jets' true-rapidity $y$ while the $\times$ marks
the pseudorapidity $\eta$.

\begin{center}
\includegraphics[width=0.39\textwidth]{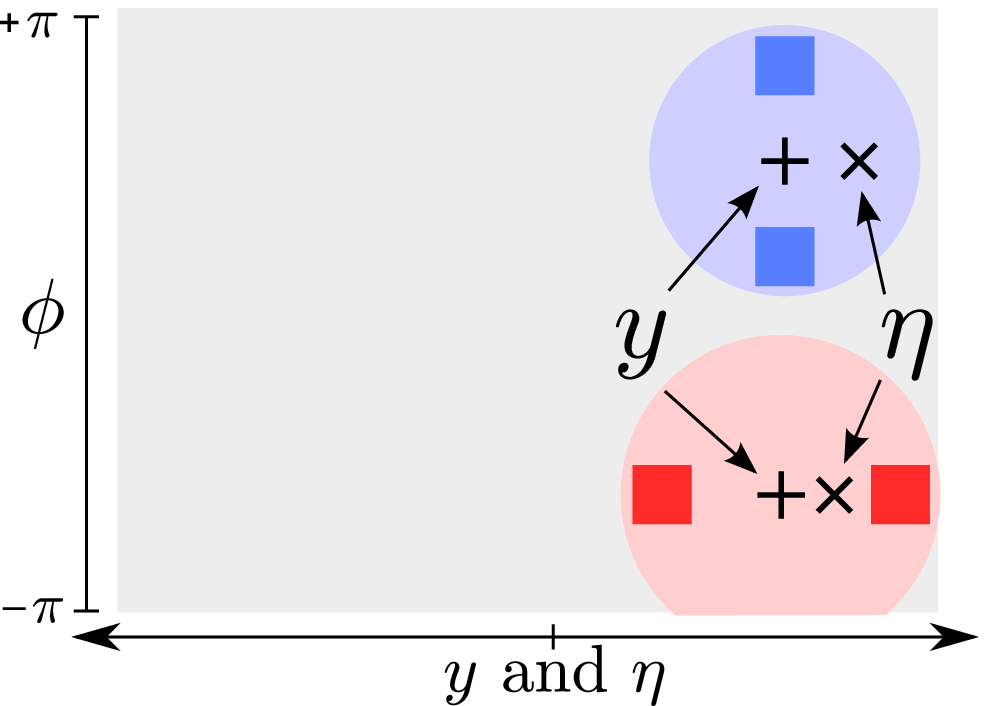}
\end{center}

\emph{Blue/Vertical/$\Delta \phi$:}
For the constituents' $\eta \! = \! 2$ and $\Delta \phi \!= \! 1$,
the true-rapidity ($y$) of the 4-vector sum is also 2,
but the pseudorapidity ($\eta$) is 2.13.
For constituents at the same rapidity,
the vector sum is at that same rapidity. However, the
pseudorapidity of the sum will \emph{always} be larger than that of the constituents.
Pseudorapidity increases toward infinity as constituents separate in $\phi$.

\emph{Red/Horizontal/$\Delta \eta$:}
When one consituent's $\eta \!=\! 1$ and the other' $\eta \!=\! 3$,
the true-rapidity ($y$) of the 4-vector sum is exactly the average 2,
but its pseudorapidity ($\eta$) is 2.4.

\emph{In General:}
For any two massless 4-vectors with same $p_T$,
you can prove that the true-rapidity ($y$) of the sum is always the
average of the constituent rapidities, even for different $\phi$
--- exactly what you ``want'' for a centroid.
On the other hand, the sum's pseudorapidity depends on the constituent $\phi$s:
If two 3-vectors from the origin poke through the unit sphere at two points,
their sum pokes through along the great circle joining those points.
Great circles are not at constant latitude or $\eta$.
They go closer to the pole and therefore have larger $\eta$s.
In general,
\begin{itemize}
\item $Rap(p_1^\mu + p_2^\mu) = (\eta_1+\eta_2)/2$ \
   when the two massless constituents have the same $p_T$ and arbitrary $\phi$.
\item $PseudoRap(p_1^\mu + p_2^\mu)$ is ugly and depends on $\phi$'s.
\end{itemize}


\subsection{Massive Jets}

If we Taylor expand pseudorapidity ($\eta$) for small $m/p_T$,
\begin{equation}
\eta = \rap + \frac{1}{2} \cos(\theta) \frac{m^2}{p_T^2} + \mathcal{O}\left( \frac{m^4}{p_T^4} \right) \ .
\end{equation}
%
For $\eta > 1$, $\cos(\theta) > 0.76$ and is rapidly approaching 1.
For QCD jets, the average jet mass-squared in the small angle limit is \cite{Salam:2009jx}
\begin{equation}
\left< m^2 \right> = C \frac{\alpha_s}{\pi} \, p_T^2 \, R^2 \, ,
\end{equation}
where $C_{gluon}=9$ and $C_{quark}=4$.
For QCD jets with $p_T=20$\,GeV, around 80\% are gluons.
Plugging in numbers, $\eta$ will be
around 0.05 larger than true-rapidity
for $R=0.7$ jets beyond rapidities of 1.
For fat jets from a hadronic top, the difference can easily be bigger than 0.1.
This is confirmed in the {\sc Pythia8} \cite{Sjostrand:2007gs} simulations below:

\vspace{1em}

\includegraphics[width=0.4\textwidth]{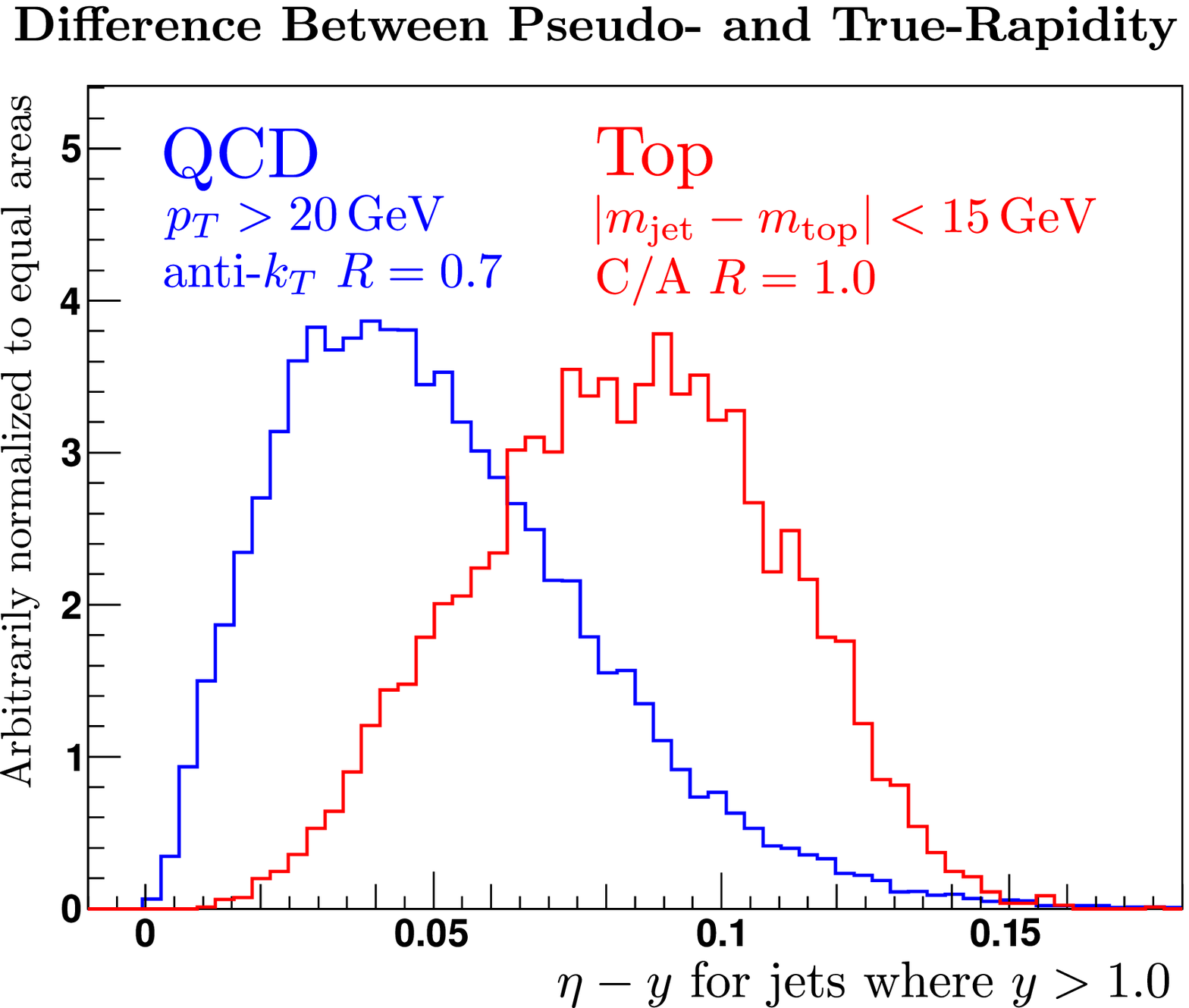}

\subsection{Centroid's $\eta$ is closer to Jet's $y$}

Jets at hadronic colliders are roughly
circular blobs in the $(\eta, \phi)$ plane of a LEGO plot.
Cone jets often define a jet's $(\eta, \phi)$ as the centroid of this blob.
But for anti-$k_T$ jets, the blob's center is at the jet's \emph{true}-rapidity.

%
%
%
%
%

The jet's $p_T$-weighted centroid is like a center-of-mass, but with $p_T$
playing the role of mass and the 2D $(\eta,\phi)$ vector playing position.
(For massless constituents, using $(y,\phi)$
would be numerically equivalent.)
For two constituents, the $\eta$-component of this centroid is defined as
\begin{equation}
\eta_\textrm{centroid} =
\frac{\eta_a \, p_T^a \, + \, \eta_b \, p_T^b }{p_T^a \, + \,  p_T^b} \ .
\end{equation}
The true-rapidity for a jet made of two massless components
(even with arbitrary $\phi$'s) turns out to be
\begin{equation}
y_\textrm{jet}  =
\frac{\eta_a + \eta_b}{2}
+ \frac{1}{2} \log \left(
   \frac{ e^{\eta_a} p_T^a + e^{\eta_b} p_T^b }
   {e^{\eta_b} p_T^a + e^{\eta_a} p_T^b}
\right) \ .
\label{eqn:jet_y_centroid}
\end{equation}
When $p_T^a = p_T^b$ as in the examples plotted, both are simply the average constituent $\eta$.
For different constituent $p_T$'s,
the centroid $\eta$ is still \emph{closer} to the jet's true-rapidity
than to the jet's $\eta$ (an ugly mess), but it's no longer exactly \emph{equal}.
These expressions \emph{are} equal
for nearby constituents until 3$^\textrm{rd}$ order,
where there is a $\Delta p_T \, \delta \eta^3$ difference.
On the other hand, the jet's $\eta$ isn't even
equivalent at first order.  This effect can be seen in the simulation below.

\begin{center}
\includegraphics[width=0.4\textwidth]{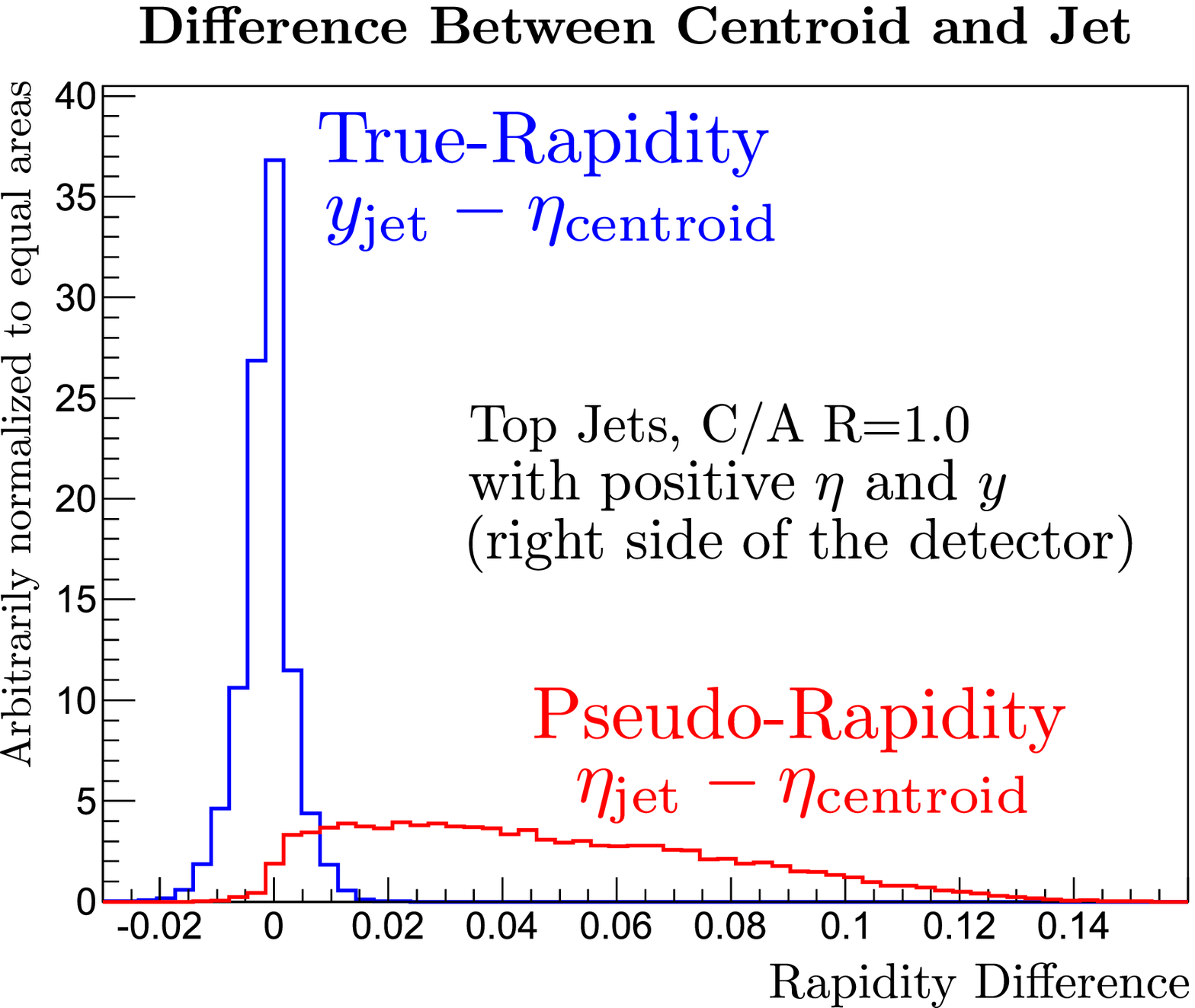}
\end{center}

An error of 0.05 to 0.1 in a jet's ``location'' doesn't sound like much,
but when calculating things like geometrical moments around the
center of a jet, using the pseudorapidity ($\eta$) \emph{systematically} gives the wrong ``center.''
Since the $p_T$ of a jet is extremely concentrated around its center,
it's important to use true-rapidity.




\subsection{Calorimeter Geometry}

You might not want jets whose edges fall off the calorimeter or tracker.
Even this geometric requirement leads to a cut on the true-rapidity of the jets, not the pseudorapidity.
Anti-$k_T$ jets of radius R=0.5 won't include additional soft constituents unless $\Delta R < 0.5$,
as measured between a possible constituent and the jet axis.
In this case, $R$ is really a maximum radius.
Fastjet calculates $\Delta R$ using the jet's true-rapidity,
so it's a radius of a circle centered around $(y,\phi)_\textrm{jet}$ not $(\eta,\phi)_\textrm{jet}$.
Say your calorimeter edge is at $\eta_\textrm{edge}$=2.5.
To keep only jets whose constituents are completely within the calorimeter,
the cut is $|\rap_\textrm{jet}| < \eta_\textrm{edge} \! - \! R$.
Assigning a track or calorimeter cell a small mass
(as discussed below) is ok too, because this only lowers its true-rapidity.
The same for cracks: since most of the $p_T$ of a jet is at its center,
jets whose true-rapidities point toward a crack are the suspicious ones.


If you were cutting on pseudorapidity because you were thinking about calorimeter geometry,
you were throwing out jets you no longer need to.


\section{Use Jet $\vec p_T$, not $\vec E_T$}

For massive jets, transverse momentum ($\vec p_T$) is the right
quantity rather than transverse energy ($\vec E_T$),
since only the former is conserved.
Expanding the ratio of these for small $m/p_T$,
\begin{eqnarray}
\frac{E_T}{p_T}
= \frac{E}{p}
&=& \sqrt{ 1 + \frac{m^2}{p^2} }  \\
&=& \sqrt{ 1 +  \sin^2(\theta) \frac{m^2}{p_T^2} } \\
&\approx& 1 + \frac{1}{2} \sin^2(\theta) \frac{m^2}{p_T^2} \ .
\end{eqnarray}

For QCD $R$=0.7 jets, 
the correction averages to 10\%.
The biggest effect occurs when the jet is purely transverse.
This is unlike before, where higher rapidity jets had a greater difference
between rapidity and pseudorapidity.
This is especially important for missing $E_T$ vs $p_T$ even though  ``MET'' is easier to pronounce than ``MpT''.


\section{Massive Constituents for Massive Jets}

Up to this point, we've treated the jet constituents as massless.
ATLAS does treat each calorimeter deposit and track as a massless
constituent when forming their jets.  CMS does this for neutrals,
but assigns pion-masses to charged particle-flow candidates.
This make little difference for the jet's $p_T$ or its $(y,\phi)$ location,
but can be a 5-20\% effect on the jet mass.

When a group of particles is boosted,
the fraction of their energy that comes from their rest mass
(rather than their momentum) can become negligible.
This is why using massless constituents makes little different to jet energy and $p_T$.
But boosting a system should keep the the invariant mass... invariant.

For two light particles of mass $m$ going back to
back, which together form a heavy invariant-mass $M$.  If each is made
massless by 
rescaling its energy, their invariant-mass-squared
is no longer $M^2$, but $M^2-4m^2$.
Boosting first and then making the particles massless
decreases their combined invariant mass by the same amount.

For $N$ such particles of mass $m$,
the altered invariant mass is $M^2 - N^2 m^2$.  For small $m/M$,
the reduction in the jet mass is approximately
\begin{equation}
\delta M
\sim -\frac{N^2}{2} \frac{m^2}{M} \ .
\label{eqn:mass_error}
\end{equation}


Another way to estimate this is by
summing constituent 4-vectors,
each made up of a massless part and a small
correction giving it mass $m_i$ to leading order.
Let them all go in the $x$-direction.
The i$^\mathrm{th}$ 4-momentum is
\begin{eqnarray}
p_i^\mu
&\sim& (E_i, E_i, 0, 0) + (0, -m_i^2 / 2 E_i, 0, 0)  \nonumber \\
&\equiv& p^\mu_{i_0} + \Delta^\mu_i
\end{eqnarray}

\

\

A jet made of $N$ of these constituents has an
energy of $E_\mathrm{jet}=\sum_i E_i$ and an invariant mass of
%
\begin{eqnarray}
M^2
&=& \left( \sum_i p^\mu_i \right)^2  
= \left( \sum_i p^\mu_{i_0} + \sum_i \Delta^\mu_i \right)^2  \nonumber \\
&\sim& \left( \sum_i p^\mu_{i_0} \right)^2 + 2 \left( \sum_i p^\mu_{i_0} \right) \cdot \left( \sum_i \Delta^\mu_i \right)  \nonumber \\
&=& M^2_0 + 2 E_\mathrm{jet} \sum_i \frac{m_i^2}{2 E_i} \nonumber \\
&\sim& M^2_0 + N^2 \langle m_i^2 \rangle
\end{eqnarray}
This gives the same fractional error as equation~(\ref{eqn:mass_error}).
For a top jet, $N$ is around a hundred,
constituents are mostly pions, kaons, and protons.
Plugging in,
\begin{equation}
 \delta M_\mathrm{top} \sim   -\frac{100^2\times 0.33^2}{2 \times 173} = -3~{\rm GeV}
\end{equation}
Below is a Monte Carlo of the difference in jet mass when constituents are made massless.

\begin{center}
\includegraphics[width=0.4\textwidth]{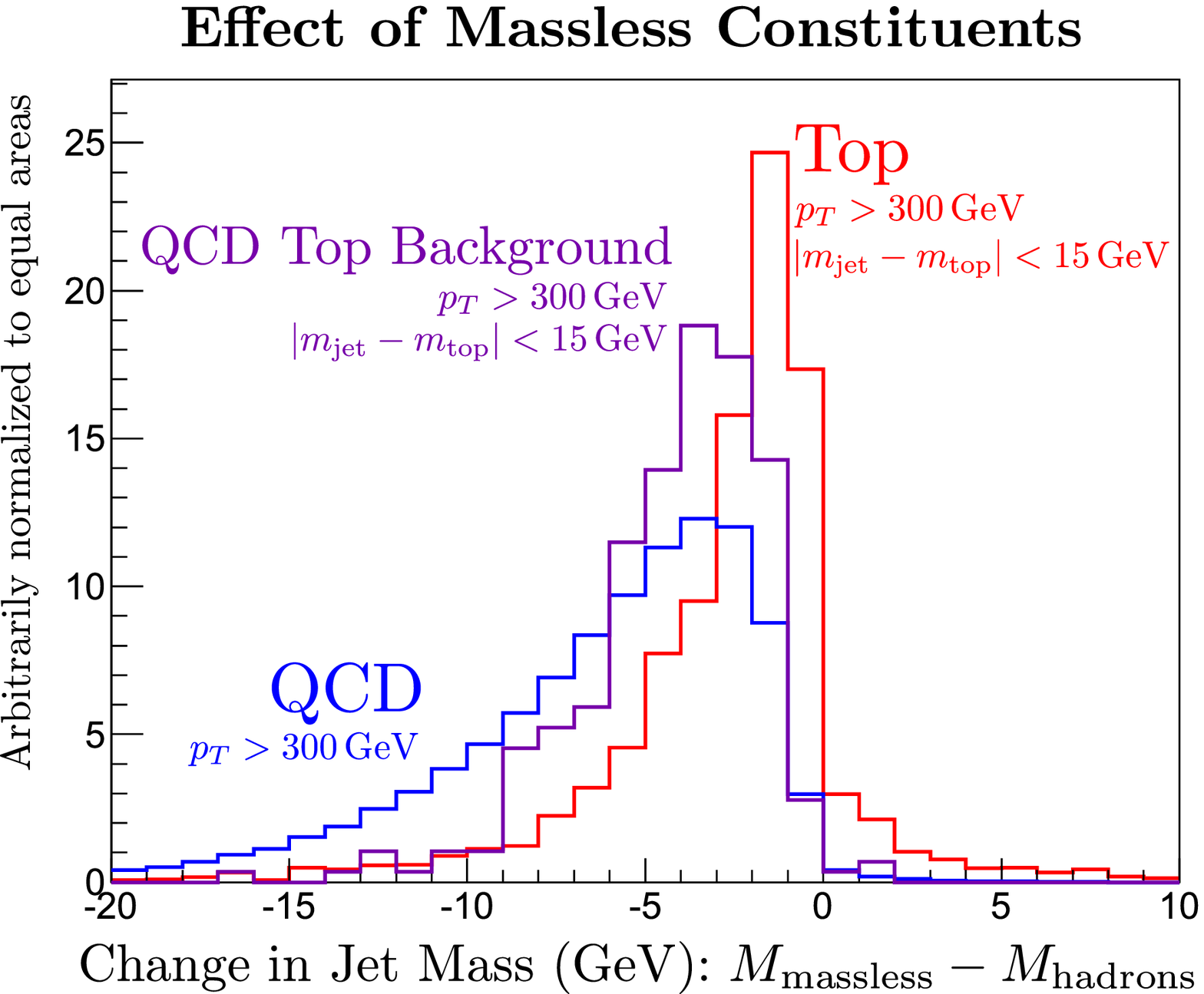}
\end{center}



All three issues addressed here lead to small changes
in observed quantities,
but these are within experimental resolution.
This note was written in 2012 and has been circulating informally. We were asked to provide something to cite. We'd like to thank Gavin Salam, Yvonne Peters, and Matthew Schwartz for helpful discussions.

\end{document}